# Prediction on Elastic Properties of Nb-doped Ni Systems

Jia Song[1], Zhibin Gao[2], Liang Zhang[1, *], Wenheng Wu[3, *], Beibei He[3], Lin Lu[3]

1 Shanghai Engineering Research Center of 3D Printing Materials, Shanghai 200437, China

2 Center for Phononics and Thermal Energy Science, China-EU Joint Center for Nanophononics, Shanghai Key Laboratory of Special Artificial Microstructure Materials and Technology, School of Physics Sciences and Engineering, Tongji University, Shanghai 200092, China

3 Shanghai Research Institute of Materials, Shanghai 200437, China

**Abstract**

On the basis of first-principles simulation, the structure, formation enthalpy, and mechanical properties (elastic constant, bulk and shear modulus and hardness) of five Nb-doped Ni systems are systematically studied. The calculated equilibrium volume increases with the Nb concentration increasing. The computational elastic constants and formation enthalpy indicate that all Nb-doped Ni systems are mechanically and thermodynamically stable in our research. The hardness of these systems also be predicted after the bulk modulus and shear modulus have been accurately calculated. The results show that the hardness increases with the Nb concentration increasing when the Nb concentration below 4.9%, beyond which the hardness will decrease within the scope of our study.

**Keywords:** First-principles simulation; Nb-doped Ni system; Formation enthalpy; Mechanical property

## 1. Introduction

Due to the excellent performances such as high temperature strength, high oxidation resistance and hot corrosion resistance, nickel-based superalloys are currently the key materials in the aerospace field of turbine blades, engine component and combustion chambers.[1-3] Usually, the nickel−based superalloys contain several refractory elements with high melting point and large atomic radii. These elements play a crucial role in improving the properties of nickel−based superalloys. Because of the large difference in electronic structure and atomic radius, different elements play different roles, such as solid solution strengthening, second phase strengthening and grain boundary strengthening.[4-6]

*Corresponding author.　Tel.: 8621-65556775-565, fax: 8621-65553475.　E-mail: lzhang0126@hotmail.com
*Corresponding author.　Tel.: 8621-65556775-442, fax: 8621-65553475.　E-mail: wwhwwh2004@126.com

In order to meet the development requirements of aerospace, it is necessary to exploit high−performance nickel−based superalloys, especially in terms of good endurance property and structural stability. Many traditional experiments have been done to research the effect of concentration of refractory elements on the nickel−based superalloys. Pröbstle et al. investigated five different Rhenium−free derivatives of CMSX−4 with varying contents of Titanium (Ti) and Tungsten (W). It was increasing found that the concentration of Ti and W resulted in more solid solution strengthener partitioning to the γ phase and significantly improved the creep strength in particular at high temperature and low stress.[7] Wang et al. used three-dimensional atom probe method to study two kinds of nickel−based superalloys with different contents of Ruthenium (Ru). Their results demonstrated that the addition of Ru element cause more Rhenium (Re) partitioning to the γ' phase and the solution strengthening for the γ phase decreasing.[8] Yeh and Tin applied an electro−thermal mechanical testing to measure the flow stresses of five nickel-based superalloys containing different contents of Re and Ru. The results showed that flow stress significantly increased with the addition of Ru or Re, and the influence of Re on the strengthening the nickel−based superalloy was more than element Ru.[4]

In recent year, duo to first-principles simulation providing accurate information on energetic and electronic structure from the atomic and electronic level, many first-principles simulations have been performed to further investigate high-temperature alloy in the aerospace field.[9-14] In this paper, we focus on the effect of strengthening element on the nickel−based superalloys. Geng et al. systematically investigated the site preference of platinum group metals added in the γ'−$Ni_3Al$ by the first−principle total energy calculations. The results suggested Mo preferred the Al site, and the other platinum group metals (Ru, Rh, Pd, Ir, Os and Pt) had a site preference for Ni site, in which the capability of element Os was weak.[15] Gong et al. applied density functional theory and Debye−Grüneisen model to study the properties of $Ni_3Al$ with the additions of single alloying elements (W, Re, Mo, Ta and Ru) and co−alloying elements (WRe, WMo, ReMo, WTa, ReTa, WRu and ReRu). It was found that single alloy elements had enhancement effect on mechanical and thermodynamic properties of $Ni_3Al$, where the effect of W and Re was similar. For the co-alloying elements, there were no distinct synergistic but simple combined strengthening effects on $Ni_3Al$.[16] Zhao et al. investigated the influence of vacancy on the site preferences of alloying elements Re, Mo, Ta and Cr at the γ/γ' interface in nickel−based superalloys by density functional theory. The calculated results showed that the most preferable substitution sites of alloying elements didn't be changed by a Ni vacancy on (001)γ' plane, except Cr element, and the alloying elements at Al site in γ' enhanced the interfacial bonding strength of γ/γ' interface, among which the strengthening effect of Re was the best.[17]

Although many solid-solution strengthening elements, such as Ru, Rh, Pd, Ir, Os, Re, Mo, Ta, Cr and Pt, have been researched by first−principles simulations, there are few studies about Nb element which can improve solid solution strengthening effect of γ matrix phase (Ni crystal). In order to reveal the influence of Nb on γ matrix phase from the atomic and electronic perspective, Ni system doping with Nb element in five different concentrations are systematically investigated by first−principles simulations. The elastic constant and modulus are firstly calculated to study the effects of the Nb content on the mechanical properties of the above five Ni systems. In addition, with the aid of Pugh's modulus ratio, the hardness of these systems will be predicted after the bulk modulus and spear modulus are accurately calculated. Finally, the Total Density of States (TDOS) and the Partial Density of States (PDOS) of Nb in all systems are investigated.

## 2. Methodology
### 2.1. Density functional theory calculations

All spin-polarized DFT calculations were performed with the Vienna Ab−Inition Simulation Package (VASP).[18-20] The exchange−correction energy of electron was included via the generalized−gradient approximation (GGA) using the Perdew−Burke−Ernzerhof (PBE) functional.[21, 22] And the ion−electron interaction was treated with projector augmented wave (PAW) approach.[23] The electronic wave functions were expanded using the plane wave base functions with a cutoff energy of 500 eV.[24] The energy convergence criterion of the electronic self−consistency was set to $10^{-5}$ eV per atom, and the force convergence criterion was $10^{-2}$ eV/Å per atom. Valence electrons used in this study were Ni ($3p^64s^23d^8$) and Nb ($4p^65s^14d^4$), respectively.

In this study, five different Nb−doped Ni systems were studied: Ni107Nb1, Ni71Nb1, Ni47Nb1, Ni31Nb1 and Ni23Nb1. These supercells were generated through enlarging a Ni cell contained 4 atom by 3×3×3, 3×3×2, 2×2×3, 2×2×2 and 3×2×1 with one of the Ni atoms substituted by Nb. Duo to the different multiples in the three directions, the Ni71Nb1 (Ni47Nb1) and Ni23Nb1 were tetragonal crystal and orthorhombic crystal, respectively. But, with same multiples in three directions supercells (Ni107Nb1 and Ni32Nb1) were still cubic crystal. The structure diagram of Ni31Nb1 was used as an example and shown in figure 1. The Brillouin−zone integration of these systems was conducted using Monkhorst−Pack grids with 2×2×2, 3×3×4, 4×4×6, 8×8×8 and 4×6×12, respectively.

Once the structural optimization of Nb−doped Ni systems, fcc Ni crystal and Nb crystals were completed, the ground state total energies of these system would be obtained by first-principles calculations.[25, 26] Then the formation enthalpy ($\Delta H(Ni_mNb_n)$) of Nb-doped Ni system could be calculated by the following formula,

$$\Delta H(Ni_mNb_n) = \frac{E_{total}(Ni_mNb_n) - mE(Ni) - nE(Nb)}{m+n} \quad (1)$$

where $E_{total}(Ni_mNb_n)$ was the total energy of an Ni$_m$Nb$_n$ system, $E(Ni)$ was the total energy of an Ni atom in the pure fcc Ni crystal and $E(Nb)$ was the total energy of an Nb atom in the pure fcc Nb crystal.

**2.2. Elastic constants**

The elastic constants determined the stiffness of a crystal against an externally applied strain. In the case of small deformation, there was a quadratic dependence of the internal energy on the strain tensor. The elastic constants described the quadratic relationship and were given by[27-31]

$$C_{ijkl} = \frac{1}{V}\left[\frac{\partial^2 E(V,\{\varepsilon_{mn}\})}{\partial_{\varepsilon_{ij}}\partial_{\varepsilon_{kl}}}\right]_{\varepsilon=0} \quad (2)$$

in which, $E(V, \{\varepsilon_{mn}\})$ was the internal energy of the crystal after strain tensor $\varepsilon_{mn}$ applied, $V$ was the volume of the unstrained crystal. Generally, the fourth−rank elastic constant $C_{ijkl}$ had no more than 21 independent components. The higher the symmetry of the crystal, the less number of independent components.

For cubic crystals Ni107Nb1 and Ni31Nb1, there were three distinct, non-vanishing elastic constants, which were $C_{11}$, $C_{12}$ and $C_{44}$. The applied strain modes were given in table 1.[32] The deformation magnitudes ε from -0.016 to 0.016 in the step of 0.004 were used in the first and second strain modes, and ε from -0.04 to 0.04 in the step of 0.01 were applied in the third strain mode. In the case of orthorhombic crystal Ni23Nb1, nine independent components of elastic constants were $C_{11}$, $C_{22}$, $C_{33}$, $C_{12}$, $C_{13}$, $C_{23}$, $C_{44}$, $C_{55}$ and $C_{66}$, respectively. The applied strain modes were listed in table 2.[32] The deformation magnitudes ε from -0.016 to 0.016 in the step of 0.004 were used in the nine strain modes. For tetragonal crystals Ni71Nb1 and Ni47Nb1, there were six independent components of elastic constants, which were $C_{11}$, $C_{33}$, $C_{12}$, $C_{13}$, $C_{44}$ and $C_{66}$. The applied strain modes were summarized in table 3.[32] The deformation magnitudes ε from -0.016 to 0.016 in the step of 0.004 were used in the six strain modes. All simulated elastic constants were obtained by cubic fitting of energy−strain curves because this method yield the smallest errors compared to quadratic fitting and quartic fitting.

Once the independent elastic constants were accurately calculated, there were two researches could be done. On the one hand, the mechanical stability of Nb−doped Ni systems could be investigated. For the cubic systems, the mechanical stability restrictions were formulated in terms of the elastic constants as follows,

$$C_{11} > 0, C_{44} > 0, C_{11} - C_{12} > 0, C_{11} + 2C_{12} > 0 \tag{3}$$

For the tetragonal systems, the mechanical stability restrictions were formulated in terms of the elastic constants as follows,

$$\begin{aligned} &C_{11} > 0, C_{33} > 0, C_{44} > 0, C_{66} > 0 \\ &C_{11} - C_{12} > 0, C_{11} + C_{33} - 2C_{13} > 0 \\ &2(C_{11} + C_{12}) + C_{33} + 4C_{13} > 0 \end{aligned} \tag{4}$$

For the orthorhombic systems, the mechanical stability restrictions were formulated in terms of the elastic constants as follows,

$$\begin{aligned} &C_{11} > 0, C_{22} > 0, C_{33} > 0, C_{44} > 0, C_{55} > 0, C_{66} > 0 \\ &C_{11} + C_{22} + C_{33} + 2(C_{12} + C_{13} + C_{23}) > 0 \\ &C_{11} + C_{22} - 2C_{12} > 0, C_{11} + C_{33} - 2C_{13} > 0 \\ &C_{22} + C_{33} - 2C_{23} > 0 \end{aligned} \tag{5}$$

On the anther hand, based on elastic constants, the shear modulus $G$ and the bulk modulus $B$ of crystal were able to be obtained by the Voigt−Reuss−Hill approximation.[33] In the case of cubic crystal, the bulk modulus and the shear modulus were given by

$$B_V = \frac{1}{3}(C_{11} + 2C_{12}) \tag{6}$$

and

$$G_V = \frac{1}{5}[(C_{11} - C_{12}) + 3C_{44}] \tag{7}$$

In the case of orthorhombic crystal, the shear modulus and the bulk modulus were given by

$$B_V = \frac{1}{9}[C_{11} + C_{22} + C_{33} + 2(C_{12} + C_{13} + C_{23})] \tag{8}$$

and

$$G_V = \frac{1}{15}[C_{11} + C_{22} + C_{33} + 3(C_{44} + C_{55} + C_{66})] \tag{9}$$

In the case of tetragonal crystal, the shear modulus and the bulk modulus were given by

$$B_V = \frac{1}{9}[2(C_{11} + C_{12}) + C_{33} + 4C_{13}] \tag{10}$$

and

$$G_V = \frac{1}{30}[4C_{11} - 2C_{12} + 2C_{33} - 4C_{13} + 12C_{44} + 6C_{66}] \qquad (11)$$

Knowing the shear modulus and the bulk modulus, the hardness of polycrystalline materials was correlated with the product of the squared Pugh's modulus ratio $k$ and the shear modulus $G$ according to the work of Chen et al..[34] The mathematical expression was written as,

$$H_V = 2(k^2 G)^{0.585} - 3 \qquad (12)$$

in which, $k$ was the Pugh's modulus ratio, namely, $k=G/B$.

## 3. Results and discussion

### 3.1. Structural optimization

The free energies of a series of Nb−doped Ni systems are calculated by first−principles simulations, and the equilibrium volumes are optimized by fitting these free energies to four−parameter Birch−Murnaghan equation of state as shown in figure 2.[35, 36] All curves fitting converge satisfactorily and the fitted equilibrium volume and lattice parameters are summarized in table 4. It can be found the calculated lattice parameters of cubic Ni crystal (consisting of 4 atoms) at 3.51 Å is good agreement with experimental value of 3.52 Å.[37] The deviation between experimental and theoretical value is less than 0.28%, so we have reason to consider that structural optimization method in this study provides satisfactory lattice parameters. In addition, from figure 2, it can be seen that the equilibrium volume per atom of Nb−doped Ni systems increases with concentrations of Nb. As the concentrations of Nb increase from 1.46% to 6.44%, the equilibrium volume per atom of these systems increases from 10.80 Å³/atom to 11.06 Å³/atom. The phenomenon can be explained from two aspects. Firstly, the atomic radius of Nb (1.85 Å) is more than that of Ni (1.35 Å), so the substitution of Nb increase the volume of Ni crystal.[38] Secondly, more Ni atoms are replaced by Nb atoms with the increase of Nb content.

According to equation (1), the formation enthalpies of five Nb−doped Ni systems are calculated and shown in Figure 3. Within the scope of our study, the values of $\Delta H(Ni_m Nb_n)$ and Nb content exhibit linear decrease relationship,

$$y = 0.64 - 0.60x \tag{13}$$

This is because that the interaction between Nb and Ni is gradually weakened with the Nb content decreasing. In addition, the formation enthalpies of examined Nb-doped Ni structures are -3.31, -2.09, -1.34, -0.86 and -0.11 KJ·mol$^{-1}$·atom$^{-1}$, which means that these compositions are stable in the ground state.

### 3.2. Elastic constants and mechanical stability

The elastic constants of Ni crystal summarized in table 5 are 260.5, 130.7 and 155.5 GPa for $C_{11}$, $C_{12}$ and $C_{44}$, respectively, consistent with experimental values of 246.5, 147.3 and 124.7 GPa.[37] These elastic constants also agree with the previous literature values at about 233 GPa for $C_{11}$, 154 GPa for $C_{12}$ and 128 GPa for $C_{44}$ calculated from embedded−atom−method functions.[37] Therefore, it is concluded that the method of calculated elastic constants in this study is reasonable and can be applied reliably to predict the elastic constants of system of which experimental measurements have not been done.

The array of elastic constants for Nb−doped Ni systems are obtained by using first-principles simulations and summarized in table 5. Firstly, it is interesting find that increase of Nb content leads to decrease of all elastic constants values and the reduction of individual elastic constants is different and nonlinear, especially for $C_{12}$. This can attributed that the elastic properties achieved by the first-principles method are extremely susceptible to the parameters and size of atomic models used in computations. What's more, based on these elastic constants, the mechanical stability of Nb−doped Ni systems can be investigated. It can be seen that the elastic constants of cubic systems (pure Ni, Ni31Nb1 and Ni107Nb1) satisfy the equation (3). The elastic constants of tetragonal systems (Ni47Nb1 and Ni71Nb1) fulfill the equation (4). The elastic constants of Ni23Nb1 systems meet the equation (5). To sum up, the five systems in our study are mechanically stable according to elastic constant analysis, which is consist with formation enthalpy results.

### 3.3. Modulus and hardness

The bulk modulus, shear modulus and hardness of Nb−doped Ni systems are calculated by using equation (6)−(12) and shown in table 4. The calculated bulk modulus for pure Ni system is 174 GPa, in good agreement with experimental value of 186 GPa.[39] In order to interpret our results, the calculated bulk modulus, shear modulus and hardness of Nb−doped Ni systems are plotted in the figure 4. As we can see, the bulk modulus and shear modulus both decrease with the increase of Nb content and can be fitted to a linear relationship,

$$y = a + bx \quad (14)$$

where, for the bulk modulus, the parameters *a* and *b* are 151.63 GPa and -1833.15 GPa; for the shear modulous, the *a* and *b* are 112.29 GPa and -859.92 GPa, respectively.

According to the intrinsic correlation between hardness and elasticity of materials, the hardness of Nb−doped Ni systems are calculated and listed in table 4. It is interesting note that the hardness of these systems increases with the concentration of Nb increasing as the Nb concentration below 4.9%, beyond which the hardness will decrease within the scope of our study. This phenomenon can be explained from saturability: when the concentration of Nb in the Ni crystal is saturated, the excess Nb element will precipitate as crystal and then do not increase the hardness of Nb−doped Ni systems.

### 3.4. Electronic structure

PDOS on d orbitals for pure Ni are presented in figure 5. Black line and red line are spin-up and spin-down electrons, respectively, which is in good agreement with former Ref. [40, 41]. Therefore, it is concluded that the computation of DOS in this study is rational and can be used to analyze the DOS of unknown system.

In order to better analyze the TDOS of different structural systems, the TDOS is divided by atomic numbers. Figure 6a shows the TDOS per average atom of the five Nb−doped Ni systems and the corresponding PDOS of Nb are portrayed in figure 6b.

Firstly, as shown in figure 6a, the peak values at -0.68 eV, -2.3 eV and 1.75 eV demonstrate that Ni and Nb have a strong hybridization which increases with Nb content increasing. And the difference in TDOS between pure Ni and Nb-doped Ni system explains the nature of vibration of formation enthalpy from atomic perspective. Secondly, the TDOS curves are all mainly located in the energy region from -7 to 10 eV, which indicates that the five compositions in our study possess excellent metallic properties. Thirdly, from the PDOS curves of Nb, it can be seen that contributions of Nb−s orbitals and Nb−p orbitals to the TDOS are almost negligible compared with Nb−d orbitals and the maximum peak of Nb−d orbitals on the right side of Fermi level change with concentration of Nb element. Comprehensively analyzing the curves of TDOS and PDOS, it can be found that the change in the TDOS may be caused by the Nb−d orbitals. In a nutshell, at the atomic level, the substitution of Nb changes the DOS of Ni crystal, thus causing macroscopic change in mechanical properties.

## 4. Conclusion

In the present work, first−principles simulations are proposed to systematically investigate the structure, formation enthalpy, elastic constants, mechanical properties (bulk modulus, shear modulus and hardness) and electronic structure of five Nb−doped Ni systems.

Firstly, the equilibrium volumes per atoms of the five systems increase with the Nb concentration increasing. Secondly, according to the elastic constants results, all the Nb−doped Ni systems are mechanically stable. The can be explained from the perspective of formation enthalpy. The calculated formation enthalpy of all systems are -3.31, -2.09, -1.34, -0.86 and -0.11 KJ·mol$^{-1}$·atom$^{-1}$, which means that these compositions are stable in the ground state. Thirdly, the hardness of all systems also be predicted after the bulk modulus and shear modulus have been precisely estimated. The results demonstrate that the hardness increases with the concentration of Nb increasing when the Nb concentration below 4.9%, beyond which the hardness will decrease within the scope of our study. Finally, we further study the electronic structure of these system, the electron DOS analysis shows that the addition of element Ni will change the TDOS to some extent.

In conclusion, the first-principles is a powerful tool that is helpful in guiding the design and optimization the composition of Ni−based superalloys.

**Acknowledgments**

This work is supported by the project to strengthen industrial development at the grass-roots level (Project Number TC160A310/19), Natural Science Foundation of Shanghai (Project Number 17ZR1409200), Shanghai Rising-star program (Project Number 18QB1400600) and Shanghai Industry Technology Institute Innovation Pioneer Program (Project Number 16CXXF006).

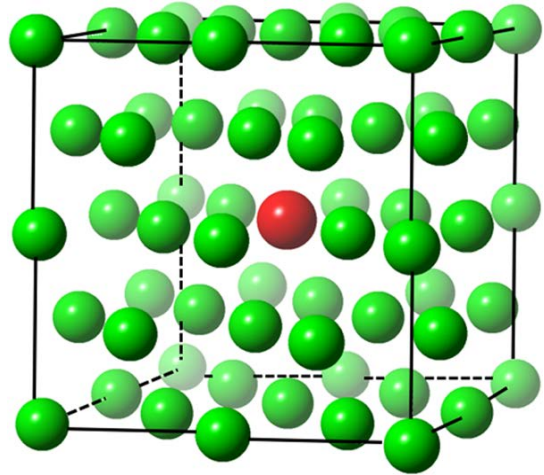

Figure 1

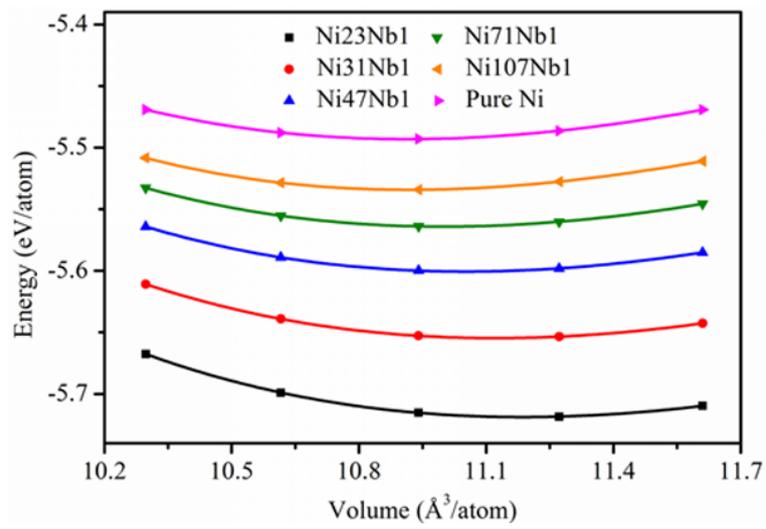

Figure 2

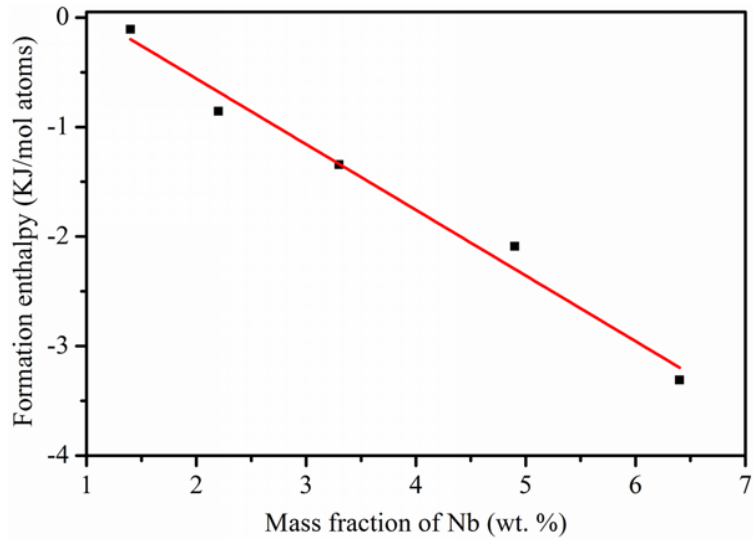

Figure 3

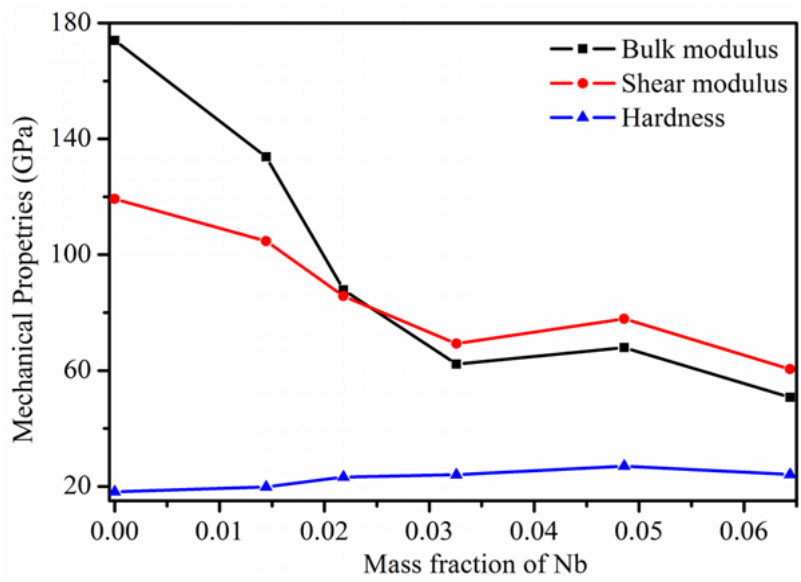

Figure 4

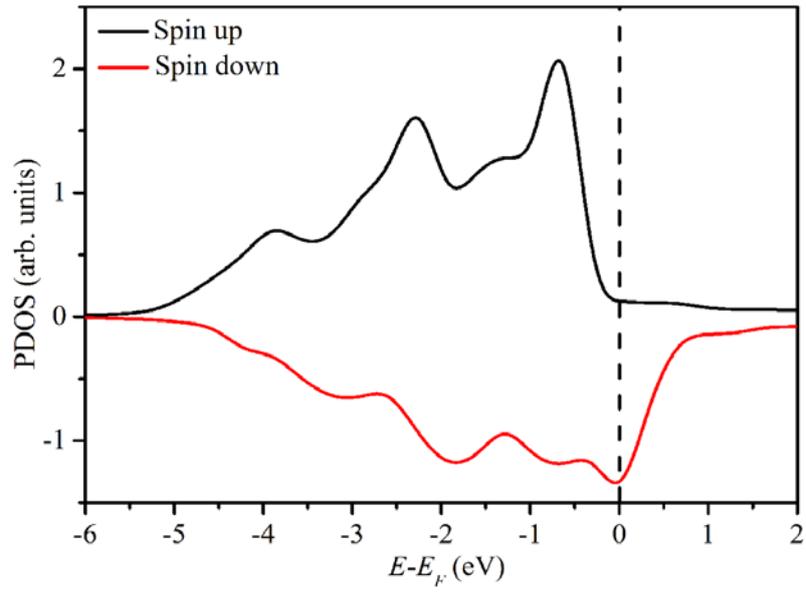

Figure 5

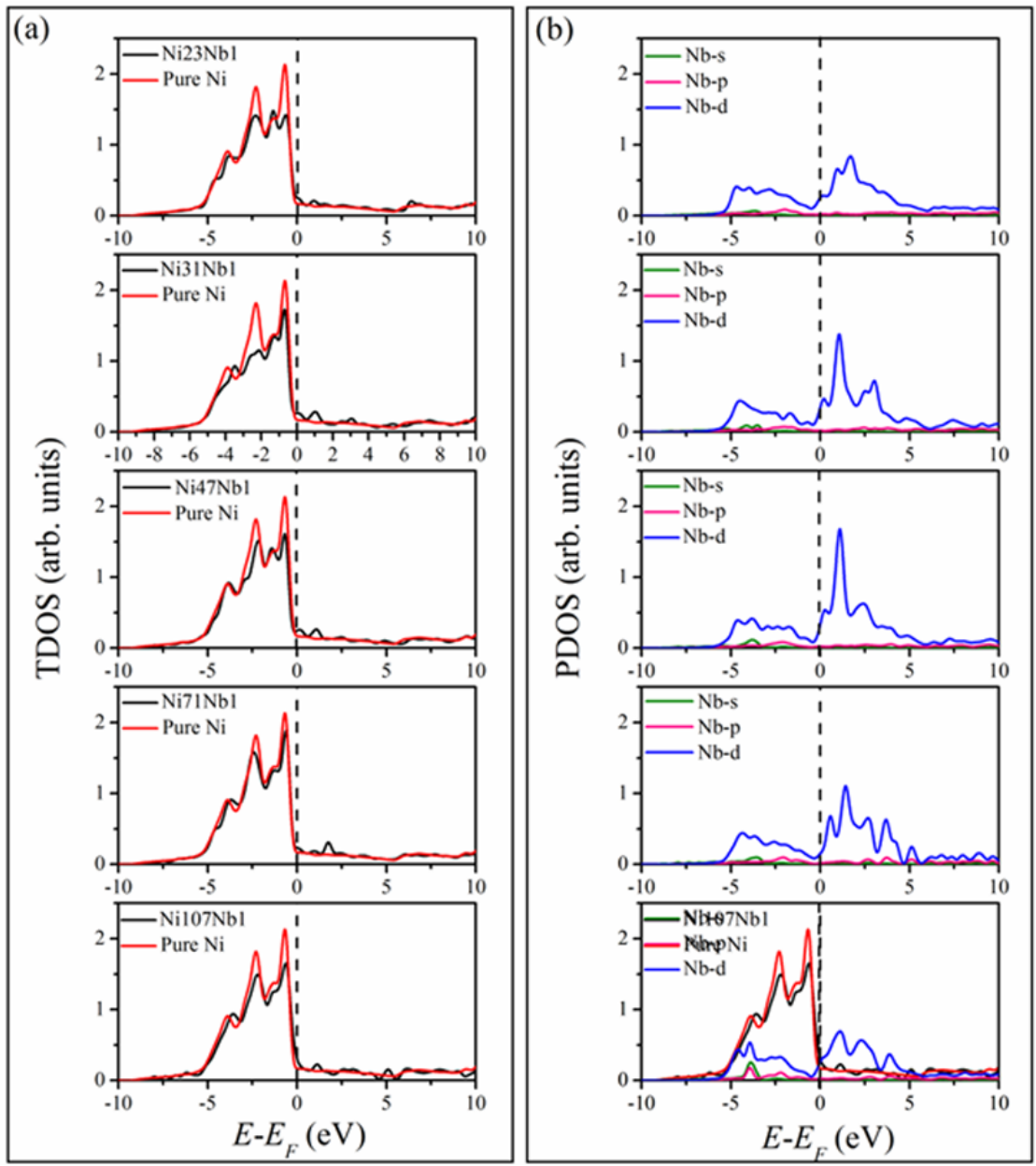

Figure 6

**Table 1. Parameterizations of the three strain modes used to calculate the three elastic constants of cubic Ni107Nb1 and Ni31Nb1.**

| Strain $I$ | Parameters (unlisted $\varepsilon_i = 0$) | $\Delta E/V_0$ to $O(\gamma^2)$ |
|---|---|---|
| 1 | $\varepsilon_1=\varepsilon_2=\gamma$, $\varepsilon_3=(1+\gamma)^{-2}-1$ | $3(C_{11}-C_{12})\gamma^2$ |
| 2 | $\varepsilon_1=\varepsilon_2=\varepsilon_3=\gamma$ | $\frac{3}{2}(C_{11}+2C_{12})\gamma^2$ |
| 3 | $\varepsilon_6=2\gamma$, $\varepsilon_3=\gamma^2(4-\gamma^2)^{-1}$ | $2C_{44}\gamma^2$ |

**Table 2. Parameterizations of the nine strain modes used to calculate the nine elastic constants of orthorhombic crystal Ni23Nb1.**

| Strain $I$ | Parameters (unlisted $\varepsilon_i = 0$) | $\Delta E/V_0$ to $O(\gamma^2)$ |
|---|---|---|
| 1 | $\varepsilon_1=\gamma$ | $\frac{1}{2}C_{11}\gamma^2$ |
| 2 | $\varepsilon_2=\gamma$ | $\frac{1}{2}C_{22}\gamma^2$ |
| 3 | $\varepsilon_3=\gamma$ | $\frac{1}{2}C_{33}\gamma^2$ |
| 4 | $\varepsilon_1=2\gamma$, $\varepsilon_2=-\gamma$, $\varepsilon_3=-\gamma$ | $\frac{1}{2}(4C_{11}-4C_{12}-4C_{13}+C_{22}+2C_{23}+C_{33})\gamma^2$ |
| 5 | $\varepsilon_1=-\gamma$, $\varepsilon_2=2\gamma$, $\varepsilon_3=-\gamma$ | $\frac{1}{2}(C_{11}-4C_{12}+2C_{13}+4C_{22}-4C_{23}+C_{33})\gamma^2$ |
| 6 | $\varepsilon_1=-\gamma$, $\varepsilon_2=-\gamma$, $\varepsilon_3=2\gamma$ | $\frac{1}{2}(C_{11}+2C_{12}-4C_{13}+C_{22}-4C_{23}+4C_{33})\gamma^2$ |
| 7 | $\varepsilon_4=2\gamma$ | $2C_{44}\gamma^2$ |
| 8 | $\varepsilon_5=2\gamma$ | $2C_{55}\gamma^2$ |
| 9 | $\varepsilon_6=2\gamma$ | $2C_{66}\gamma^2$ |

**Table 3.** Parameterizations of the six strain modes used to calculate the six elastic constants of tetragonal crystals Ni71Nb1 and Ni47Nb1.

| Strain $I$ | Parameters (unlisted $\varepsilon_i = 0$) | $\Delta E/V_0$ to $O(\gamma^2)$ |
|---|---|---|
| 1 | $\varepsilon_1=2\gamma$, $\varepsilon_2=-\gamma$, $\varepsilon_3=-\gamma$ | $\frac{1}{2}(5C_{11} - 4C_{12} - 2C_{13} + C_{33})\gamma^2$ |
| 2 | $\varepsilon_1=\gamma$, $\varepsilon_2=-2\gamma$, $\varepsilon_3=\gamma$, $\varepsilon_6=2\gamma$ | $(C_{11} + C_{12} - 4C_{13} + 2C_{33} + 2C_{66})\gamma^2$ |
| 3 | $\varepsilon_1=-\gamma$, $\varepsilon_2=-\gamma$, $\varepsilon_3=2\gamma$ | $(C_{11} + C_{12} - 4C_{13} + 2C_{33})\gamma^2$ |
| 4 | $\varepsilon_1=\gamma$ | $\frac{1}{2}C_{11}\gamma^2$ |
| 5 | $\varepsilon_3=\gamma$ | $\frac{1}{2}C_{33}\gamma^2$ |
| 6 | $\varepsilon_6=2\gamma$ | $2C_{66}\gamma^2$ |

**Table 4.** The lattice parameters, equilibrium volume per atom, bulk modulus, shear modulus and hardness of Nb-doped Ni calculated from first-principles.

| System | a Å | b Å | c Å | $V_0$ Å$^3$/atom | $B_V$ GPa | $G_V$ GPa | $H_V$ GPa |
|---|---|---|---|---|---|---|---|
| Ni23Nb1 (Cal.) | 10.60 | 7.08 | 3.54 | 11.06 | 50.7 | 60.48 | 24.1 |
| Ni31Nb1 (Cal.) | 7.06 | 7.06 | 7.06 | 10.99 | 67.9 | 77.9 | 27.0 |
| Ni47Nb1 (Cal.) | 7.04 | 7.04 | 10.55 | 10.91 | 62.2 | 69.3 | 24.0 |
| Ni71Nb1 (Cal.) | 10.54 | 10.54 | 7.04 | 10.86 | 87.9 | 85.7 | 23.2 |
| Ni107Nb1 (Cal.) | 10.53 | 10.53 | 10.53 | 10.80 | 133.8 | 104.7 | 19.8 |
| Pure Ni (Cal.) | 3.51 | 3.51 | 3.51 | 10.79 | 174.0 | 119.3 | 18.1 |
| Pure Ni (Exp.) | 3.52 | 3.52 | 3.52 | 10.90 | 186.0 | | |

**Table 5.** The elastic constants $C_{ij}$ (unit: GPa) for Nb-doped Ni systems calculated from first-principles.

| System | $C_{11}$ | $C_{22}$ | $C_{33}$ | $C_{12}$ | $C_{13}$ | $C_{23}$ | $C_{44}$ | $C_{55}$ | $C_{66}$ |
|---|---|---|---|---|---|---|---|---|---|
| Ni23Nb1 (Cal.) | 96.6 | 99.8 | 98.0 | 26.0 | 28.9 | 25.9 | 107.2 | 97.4 | 26.6 |
| Ni31Nb1 (Cal.) | 90.2 | | | 56.7 | | | 118.7 | | |
| Ni47Nb1 (Cal.) | 103.3 | | 93.0 | 43.4 | | 32.1 | 123.2 | | 35.9 |
| Ni71Nb1 (Cal.) | 169.4 | | 155.2 | 49.5 | | 40.9 | 131.0 | | 45.6 |
| Ni107Nb1 (Cal.) | 221.9 | | | 99.8 | | | 140.4 | | |
| Pure Ni (Cal.) | 260.5 | | | 130.7 | | | 155.5 | | |
| Pure Ni (Cal.)* | 233.0 | | | 154.0 | | | 128.0 | | |
| Pure Ni (Exp.) | 246.5 | | | 147.3 | | | 124.7 | | |

* data from previous embedded-atom-method simulations